\documentclass{aa}

\usepackage{graphicx}

\thesaurus{07 % A&A Section 7: Solar System
          (07.13.1) % Meteors, meteoroids.
          (02.16.1) % Plasmas.
          (02.19.2)}% Scattering.

\begin{document}

\title{On the meteor height from forward scatter radio observations}

\author{A. Carbognani\inst{1}, M. De Meyere\inst{2}, L. Foschini\inst{3} 
\and C. Steyaert\inst{4}}

\offprints{A. Carbognani}

\institute{Department of Physics, University of Parma and Istituto Nazionale
per la Fisica della Materia, Parco Area delle Scienze 7, I--43100, Parma 
(Italy); (email: albino@fis.unipr.it)
\and Radio Amateur ON4NU, Belgium 
\and Institute TeSRE -- CNR, Via Gobetti 101, I--40129, Bologna 
(Italy); (email: foschini@tesre.bo.cnr.it)
\and VVS Astronomical Association, Belgium}

\date{Received 21 April 2000; accepted 22 June 2000}

\titlerunning{On the meteor height from forward scatter radio observations} 
\authorrunning{A. Carbognani et al.}

\maketitle

\begin{abstract} It is known from theory that, by means of a plasma
physics approach, it is possible to obtain a simple formula to calculate
the approximate height of a meteor (Foschini, 1999). This formula can be
used in case of forward scatter of radio waves and has the advantage that
it does not depend on the diffusion coefficient. On the other hand, it is
possible to apply the formula to a particular type of meteor only
(overdense meteor type I), which is a small fraction of the total number
observed.  We have carried out a statistical analysis of several radio
echoes from meteor showers recorded during last years by a radio observer
located in Belgium. Results are compared and discussed with those obtained
with other methods and available in literature. 

\end{abstract}

\section{Introduction}
A meteoroid enters the Earth's atmosphere at hypersonic speed and it 
collides with air molecules. The high kinetic energy involved in the 
process determine the transformation of a solid body into a plasma, 
which can scatter radio waves and can emit light (meteor). 

During sixties and seventies several works 
investigated the formation and evolution of the meteor, with 
particular attention to diffusion, in order to study mesospheric 
winds.  A complete review of standard meteor science can be found 
in Ceplecha et al.  (\cite{CEPLECHA1}).  However, there are still some 
aspects not well understood about the physical properties of a meteor, 
specifically whether it is an ionized gas or a plasma.  During past 
years, these two appellatives were often used as synonymous in meteor 
physics, even though they indicate two different states of the matter.  
In some studies, such as those about diffusion, specific plasma 
properties are taken into account (e.g.  ambipolar diffusion); however 
in other studies, such as about radiowave scattering, the meteor is simply 
considered a long narrow column of ionized gas. 

This can appear as a futile debate, but it hides important concepts. 
Specifically, a plasma has collective properties (e.g. Langmuir 
frequency)  that an ionized gas has not. 

A first attempt to study the meteor as a plasma was carried out by 
Herlofson (\cite{HERLOFSON}).  He investigated the proper oscillations 
in the meteor and their interaction with radio waves.  But, at our 
knowledge, none continued his studies.  Only in 1999 the question of 
collective oscillations in meteoric plasma was reprised (Foschini 
\cite{FOSCHINI1}).  Perhaps, this gap may be explained by taking into 
account that, according to purposes of meteor astronomy, it was 
sufficient to use the approximation of the long narrow cylinder.

However, the meteoric plasma is something more complex than a 
reflecting rod and it is necessary to study it. There are several 
types of oscillations and instabilities, which can interact with radio 
waves. The scattering is not the only process: for example, 
fluctuations from equilibrium may lead to transformation of waves 
(longitudinal to transverse and vice versa). The question is: are such 
processes present in a meteoric plasma? 

We think that the study of plasma collective oscillations may give new 
useful tools to understand the physics of meteors.  Some basic 
concepts about meteoric plasma were settled in a previous paper 
(Foschini 1999), thereafter called Paper I. According to the theory 
exposed there, radio echoes can be divided into two classes and two 
subclasses.  Then we have underdense and overdense echoes, according 
to whether the Langmuir frequency is higher or lower than the radio 
wave frequency. Overdense echoes totally reflect electromagnetic 
waves, but the presence of binary collisions among ions and electrons 
weaken the collective oscillations of the plasma, allowing the 
propagation of the waves, even though with strong attenuation.  
Therefore, we can divide the overdense echoes into two subclasses: 
type I, when there is total reflection; type II, when binary 
collisions allow the propagation. The division between overdense type 
I and II depends on the electron--ion collision frequency, which in 
turn depends on electron density and ion cross section. In Paper 
I, for the sake of simplicity, we considered potassium ion, that 
is the chemical element with lower ionization energy. In addition, recent 
studies show that potassium seems to be much more important in the evolution 
of meteor than previously thought (von Zahn et al. \cite{MURAD}). 
With this assuption, the division between overdense echoes occurs at about 
$10^{17}$~m$^{-3}$. It is worth noting that this border can be moved 
by considering other elements. But the calculation of particle 
distribution and evolution in a meteoric plasma will be object of 
other papers. 

\begin{figure}[t]
\centering
\includegraphics[scale=0.4]{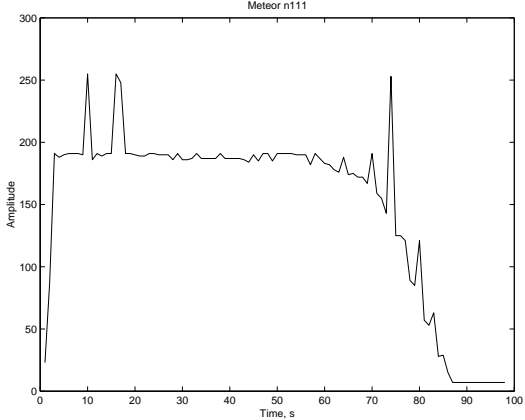}
\caption{Example of overdense type I echo.}
\label{FIG1}
\end{figure}

The overdense type I echoes derives from total reflection of radio 
waves (see Fig.~\ref{FIG1} for an example).  This allows to calculate
the height of the meteor in an easy way, as shown in the Paper I.  
Here we want to present a statistical sample of several meteor 
showers, for which we have calculated the height.  Data will be 
discussed and compared with available data in literature.

\section{A simple formula for meteor height}
We shortly recall how to calculate the meteor height, as described in
Paper I. First, we have to take into account that the plasma has 
not a definite boundary and then, the incident electromagnetic wave 
penetrates a little into it before reaching the density 
necessary to allow total reflection. We can consider this something 
similar to skin effect in metals. 

We can consider a simple geometry, as shown in Fig.~6 of Paper I, and 
then use the definition of the attenuation $a$ in decibel units:

\begin{equation}
	a=10\cdot \log \frac{|\vec{E_{i}}|^{2}} {|\vec{E_{r}}|^{2}} \ 
	[{\rm dB}]
	\label{e:decib}
\end{equation}

\noindent where subscripts $i$ and $r$ stand for incident and reflected wave. 

>From the solution of Maxwell's equations we obtain 
that, for overdense meteors type I, the electric field is:

\begin{equation}
	\vec{E}=\vec{E_{0}}e^{i(\vec{k}\cdot\vec{r}- \omega t)}
	\label{e:solu}
\end{equation}

\noindent where the wave vector in Eq.~(\ref{e:solu}) has the form:

\begin{equation}
	\vec{k}=\vec{\beta} +i\vec{\alpha}
	\label{e:wv}
\end{equation}

\noindent We refer to Paper I for explanation of symbols, even though 
they are commonly used in literature about electromagnetic fields.

We substitute Eqs.~(\ref{e:solu}) and (\ref{e:wv}) in 
Eq.~(\ref{e:decib}) and, taking into account that the amplitude of a 
totally reflected wave is equal to the amplitude of the incident wave, 
we can obtain an attenuation value of about $a=-20\alpha l\log e$, where 
$l$ is the path of the wave into the plasma:

 \begin{equation}
	l=\frac{2\delta}{\cos\phi}
	\label{e:path}
 \end{equation}

\noindent where $\phi$ is the incidence angle and $\delta = 1/\alpha$ is the 
penetration depth.  Then, Eq.~(\ref{e:decib}) becomes:

\begin{equation}
	a=\frac{-40\log e}{\cos\phi}\cong \frac{-17.36} {\cos\phi}=-17.36 
	\sec\phi \ [{\rm dB}]
	\label{e:decib1}
\end{equation}

In the case of overdense type I (total reflection) the attenuation is 
simply a function of the angle of incidence.

The Eq.~(\ref{e:decib1}) refers to an idealized case. When we deal 
with real meteors and radio waves, we have to take into account of
several factors, i.e. antenna gains, losses in radio receiver and 
trasmitter, atmospheric absorption, and the distance of reflecting 
point from trasmitter and receiver. Strictly speaking, 
Eq.~(\ref{e:decib1}) can be considered as the ``meteor cross section'' 
in the radar equation.

We can consider common factors in radar theory, as described in 
Kingsley \& Quegan (\cite{KINGSLEY}).  By means of commonly used 
values for forward scatter radar, we obtain that the attenuation 
recorded with our receiver is:

\begin{equation}
	a=20\log V_{R} - 2\ [{\rm dB}]
	\label{e:meyere}
\end{equation}

\noindent where $V_{R}$ is the received signal amplitude [V]. 
>From the amplitude of the reflected wave, we can calculate the 
incidence angle with the Eq.~(\ref{e:decib1}). 

\begin{figure}[t]
\centering
\includegraphics[angle=270, scale=0.3]{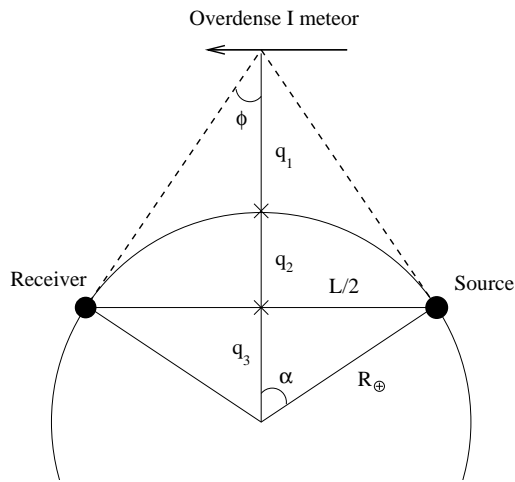}
\caption{Forward--scatter reflection geometry and height calculation
(plot not in scale)}
\label{FIG2}
\end{figure}

Therefore, we can calculate the meteor height by considering 
the geometry of Fig.~\ref{FIG2}. We can see that:

\begin{equation}
q_{1}=\frac{L/2}{\tan\phi}-q_{2}
\label{e:hei1}
\end{equation}

Taking into account that the Earth's mean radius $R_{\oplus}$ is much
larger than $L$, we can calculate $q_{2}$: 

\begin{equation}
q_{2}=R_{\oplus}-q_{3}=R_{\oplus}-\sqrt{R_{\oplus}^{2}-(L/2)^{2}}\approx
\frac{L^{2}}{8R_{\oplus}}
\label{e:hei2}
\end{equation}

By substituting Eq.~(\ref{e:hei2}) in Eq.~(\ref{e:hei1}), we obtain the
height of the reflection for an overdense meteor type I: 

\begin{equation}
	q_{1}\approx \frac{L}{2}\left(\frac{1}{\tan 
	\phi}-\frac{L}{4R_{\oplus}}\right)
	\label{e:height}
\end{equation}

\noindent where $L=2R_{\oplus}\sin\alpha$. The angle $\alpha$ is the half
angular distance between receiver and transmitter.

The distance $L/2$ plays an important role in the derivation of
the above formula. Indeed, from a geometric point of view, the specular
reflection in a forward scatter system occurs when the meteor trail lies
along a tangent to an ellipsoidal surface, with the transmitter and the
receiver stations in foci (Forsyth \& Vogan \cite{FORSCAT}). This
condition is fulfilled by different values of distances of the reflecting
point from the source and from the receiver. If we do not know the
path source--meteor--receiver, this introduces an uncertainty of
about 40~km in the height of the reflecting point (for our system). 

One way to overcome this problem is to set up a third station, but if it
is not possible, as in our case, we can reduce uncertainties by making
heuristic considerations. Indeed, as explained by Forsyth \& Vogan
(\cite{FORSCAT}), the forward scatter system is most sensitive to meteor
trails which are nearly horizontal and directed along the transmission
line. A meteor perpendicular to the source--receiver line gives an echo
that is about five times lower than the case of parallel direction
(Forsyth \& Vogan \cite{FORSCAT}).  Therefore, the choice of the
reflecting point located closely to the middle of the transmission path
appears to be reasonable and, as we shall see, is justified by facts.

\section{Observations}
The general principle of meteor observation by forward scattering of
radio waves is the following: a VHF radio receiver (30--100~MHz) is located
at a large distance (about 600--2000~km) from a transmitter at the same 
frequency. Direct radio communication is not possible, owing to the Earth's 
curvature, but the meteor allows the commuication over the horizon, by 
reflecting the transmitted signal.

In this study, observations were carried out mainly by M.~de Meyere. His 
radio receiver is located in Deurle, Belgium (longitude $3^{\circ} 37'$~E, 
latitude $51^{\circ} 00'$~N), while the trasmitter is located in Sofia 
(Bulgary). It transmits radio signals at 66.50~MHz all over the day 
with 10~kW power. In this case, values of $L$ and $\alpha$ are, respectively, 
$1751$~km and $7.9^{\circ}$.

The receiving station consists of a crossed Yagi antenna (4 elements), that 
is linked to a computer with a digital acquisition  interface (150 samples 
per second, 8 bit resolution). Data are recorded and stored into a file.

\begin{table}
\centering
\begin{tabular}{lcc}
\hline
Shower & Years & Over. I meteor \\
\hline
Geminids & 96-97 & 26 (1.1\%) \\
Leonids  &94-95-96-97-99& 44 (0.6\%) \\
Lyrids   &95-96-97 & 41 (1.9\%) \\
Quadrantids & 95-98 & 20 (1.2\%) \\
\hline
\end{tabular}
\caption{Observed meteor showers. In the last column, the percentage 
indicates the number of overdense meteors type I compared with the total 
number of recorded meteors.}
\label{tab1}
\end{table}

\begin{table}
\centering
\begin{tabular}{lcc}
\hline
Shower & $V$, km/s & $\bar{q}_{1}\pm\sigma$, km \\
\hline
Geminids & 36 & $101\pm 4$ \\
Leonids & 72 & $101\pm 8$ \\
Lyrids   & 47 & $101\pm 6$ \\
Quadrantids & 43 & $ 103\pm 3$ \\
\hline
\end{tabular}
\caption{Mean speed and height of overdense I meteor from several meteor 
showers.}
\label{tab2}
\end{table}

\begin{figure}[t]
\centering
\includegraphics[angle=0, scale=0.4]{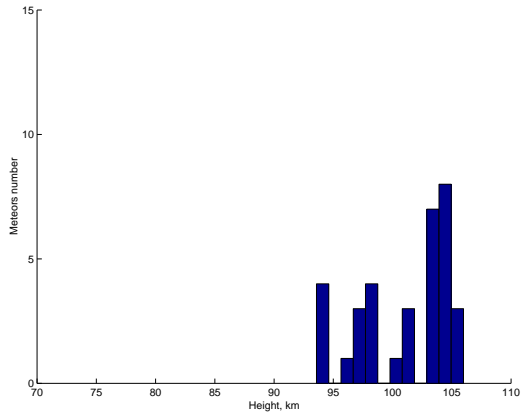}
\caption{Height distribution for Geminids meteor shower.}
\label{FIG3}
\end{figure}

\begin{figure}[t]
\centering
\includegraphics[angle=0, scale=0.4]{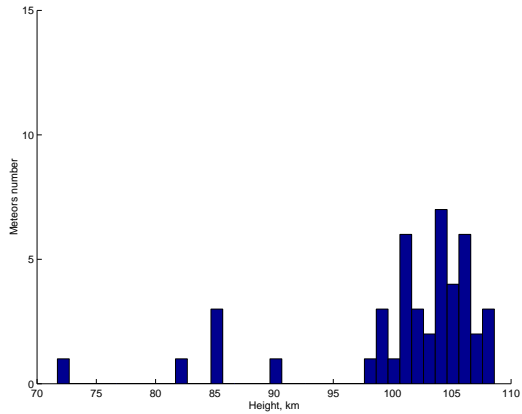}
\caption{Height distribution for Leonids meteor shower.}
\label{FIG4}
\end{figure}

\begin{figure}[t]
\centering
\includegraphics[angle=0, scale=0.4]{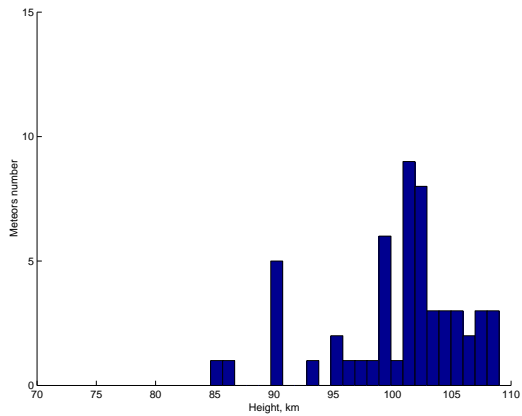}
\caption{Height distribution for Lyrids meteor shower.}
\label{FIG5}
\end{figure}

\begin{figure}[t]
\centering
\includegraphics[angle=0, scale=0.4]{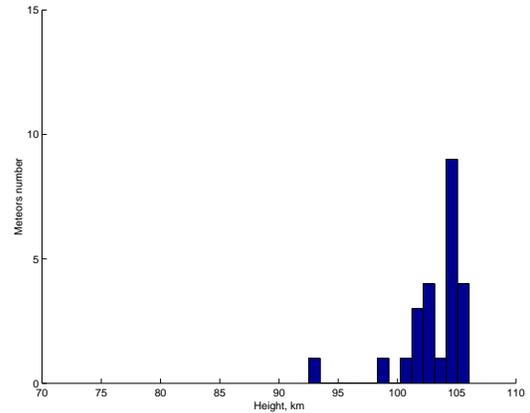}
\caption{Height distribution for Quadrantids meteor shower.}
\label{FIG6}
\end{figure}

Observations of several meteor shower were carried out during several years 
(see Table~\ref{tab1}), with a total number of meteors recorded equal to 
13401. The overdense I meteors are 131 (about 1\% of the total) and the 
height distribution for analysed showers are shown in 
Figs.~\ref{FIG3}--\ref{FIG6} (each bin is 1~km wide). 

Values in Table~\ref{tab1} refer to all echoes recorded during
shower days. In order to evaluate also the sporadic background, we have
analysed some days in February, without any shower. We have found a mean
value for sporadic overdense type I meteors of about 0.2 meteor per hour, so
that the contribution of background can be considered negligible.

Measured heights are in the range between 70 and 110~km, in good agreement 
with typical meteor heights (60--110 km), even though the large part of
meteors are in the range 95--110 km. The peaks of distributions are not
centered, but are located toward right. However, this seems to be an
effect dues to a low number of data. Indeed, the best fit for the
observations, calculated with the $\chi^2$ test, is a gaussian
distribution (therefore $\sigma$ is calculated with standard methods
for this type of distribution). In Table~\ref{tab2} values of mean speed
(Allen \cite{ALLEN}) and height of overdense I meteor are shown. 

\section{Analysis and discussion}
Data obtained here show that the mean height is independent from the entry 
speed of the meteoroids (see Table~\ref{tab2}). On the other hand, it is 
known that the height depends on entry speed of meteoroids: for example, 
Greenhow \& Lovell (\cite{GREENHOW}) wrote that the highest speed sporadic 
meteors, moving at 60--70~km/s, ionize at a mean height of 100~km, whereas 
those with minimum speed (11.2~km/s) reach a mean height of 85~km. 

The theory of radio meteor height was elaborated by Kaiser 
(\cite{KAISER1}, b) and recently Belkovich et al. (\cite{BELKOVICH}) proposed 
some changes, in order to take into account the fragmentation. Kaiser found 
that the width of the height distribution depends on the atmospheric scale 
height and the mass distribution of incoming meteoroids. The mean height 
depends strongly on meteor speed, through two coefficients named $k_{1}$ and 
$k_{2}$, and depends also on the probability of ionization. Kaiser's theory 
refers to the point of maximum ionization, but it is known that in 
experimental radio observations the height of reflecting point does not 
necessarily lie in the point of maximum ionization 
(Greenhow \& Lovell \cite{GREENHOW}).

If we observe a meteor shower with a given mean speed and mass
distribution, the height distribution is related to the length of
ionization curve. The point of maximum ionization corresponds to the most
probable height. It is worth noting that Kaiser's theory refers to
underdense meteors. Only McKinley (\cite{MCKINLEY}) referred to
overdense meteors and found no clear dependence on speed.  Moreover, he
found a two peaks distribution: the main peak is located at about 95~km, and 
the second one at 106~km.

We can try to explain the differences between our results and data available 
in literature. The first reason is that we try to analyse overdense meteors, 
while the large part of published data refer to underdense meteors. It is 
very interesting to note that for bright meteors, which are surely overdense, 
the radio height of maximum echo duration is well above the mean height of 
maximum light, obtained from photographic data (Millman \& McKinley 
\cite{MILLMAN}). On the other hand, the situation is reversed for faint 
meteors.  

The reason for this difference is that overdense type I meteors reflect 
totally the incoming electromagnetic wave. Total reflection is allowed only 
when plasma frequency is higher than the radio frequency and the collision 
frequency in the plasma is negligible (see Paper I). These conditions are 
achieved indipendently from speed of incoming of meteoroid, but it depends on 
the mass and chemical composition of the body. Once an overdense type I meteor 
is created, the signal amplitude of the reflected wave remains constant until 
the collision frequency in the plasma or recombination and attachment 
processes subtract energy to the plasma frequency. We can say that collective 
properties of the plasma, which generate the long plateau of overdense type 
I meteor, ``hide''  in some way some properties of the incoming meteoroid.

Concerning the two peaks found by McKinley (\cite{MCKINLEY}), we
note that our distributions show only the secondary peak. This can be
explained by taking into account that while McKinley made no distinction
between overdense meteors, we have considered only overdense type I
meteors.

\section{Concluding remarks}
We have carried out an analysis of several overdense radio echoes, 
recorded during last years by a radio observer located in Belgium. 
We have analysed a particular class of overdense meteors (type I) 
and measured height distributions are in good agreement with previous results
obtained by McKinley (\cite{MCKINLEY}), even though only for the secondary
peak. We suppose that the first peak in McKinley's work should be due to
overdense type II meteors, while the secondary peak, recorded also by our
system, appeared to be due to overdense type I meteors.

We think that collective properties of the meteoric plasma (Langmuir
oscillations) hide some characteristics of the originary cosmic body,
specifically there is no clear dependence on speed.  Further study, mainly
theoretical and able to take into account collective properties of plasma,
are required to assess the particle dynamics in the meteor.

It should be noted that our studies were carried out with an amateur
forward scatter system and we have no full control on it. Moreover,
heuristic considerations were introduced in order to minimize
uncertainties, but results showed that they were justified by facts.  
The agreement with previous works, with other techniques, supports our
conclusions. The future availability of a full forward scatter system would
be of great help in more detailed studies.

\begin{acknowledgements} 
LF wishes to thank P.~Farinella for useful
discussions. Authors are also grateful to the referee, P.~Pecina,
for useful comments. This research has made use of \emph{NASA's
Astrophysics Data System Abstract Service}.  
\end{acknowledgements}

\end{document}